\documentclass[11pt,a4paper]{article}

\usepackage{epsfig}
\usepackage{amsmath}
\usepackage{amssymb}
\usepackage{verbatim}
\usepackage{bm}
\usepackage{dsfont}
\usepackage[footnotesize]{caption}
\usepackage{subfigure}
\usepackage{cite}
\usepackage{textcomp}
\usepackage{calc}
\usepackage{geometry}
\usepackage{bbm}
\usepackage{xcolor} 
\usepackage{multirow}
\usepackage{color}
\usepackage{float}
\usepackage{bbold}

\usepackage{hyperref}
\hypersetup{colorlinks=true,urlcolor=blue,linkcolor=red,citecolor=green!60!black}

\begin{document}


\noindent December 2016
\hfill IPMU 16-0214
 
\vskip 2.5cm

\begin{center}
{\LARGE\bf Minimal Seesaw Model with a Discrete\\\smallskip Heavy-Neutrino Exchange Symmetry}

\vskip 2cm

\renewcommand*{\thefootnote}{\fnsymbol{footnote}}

{\large
Thomas~Rink,$^a$
Kai~Schmitz,$^{a,\,\hspace{-0.25mm}}$%
\footnote{Corresponding author. E-mail: kai.schmitz@mpi-hd.mpg.de}
and Tsutomu~T.~Yanagida\,$^b$}\\[3mm]
{\it{
$^{a}$ Max Planck Institute for Nuclear Physics (MPIK), 69117 Heidelberg, Germany\\
$^{b}$ Kavli IPMU (WPI), UTIAS, The University of Tokyo, Kashiwa, Chiba 277-8583, Japan}}

\end{center}

\vskip 1cm

\renewcommand*{\thefootnote}{\arabic{footnote}}
\setcounter{footnote}{0}


\begin{abstract}


\noindent We present a Froggatt-Nielsen flavor model that yields a minimal
realization of the type-I seesaw mechanism.
This seesaw model is minimal for three reasons:
(i) It features only two rather than three right-handed sterile neutrinos: $N_1$ and $N_2$,
which form a pair of pseudo-Dirac neutrinos;
(ii) the neutrino Yukawa matrix exhibits flavor alignment, i.e., modulo small perturbations,
it contains only three independent parameters; and 
(iii) the $N_{1,2}$ coupling to the electron flavor is parametrically
suppressed compared to the muon and tau flavors, i.e., the neutrino Yukawa matrix
exhibits an approximate two-zero texture.
Crucial ingredients of our model are (a) Froggatt-Nielsen flavor charges
consistent with the charged-lepton masses as well as (b) an approximate,
discrete exchange symmetry that manifests itself as
$N_1 \leftrightarrow i\, N_2$ in the heavy-neutrino Yukawa interactions
and as $N_1 \leftrightarrow N_2$ in the heavy-neutrino mass terms.
This model predicts a normal light-neutrino mass hierarchy,
a close-to-maximal $CP$-violating phase in the lepton mixing matrix,
$\delta \simeq 3/2\,\pi$, as well as resonant leptogenesis in accord
with arguments from naturalness, vacuum stability, perturbativity
and lepton flavor violation.


\end{abstract}


\thispagestyle{empty}
 
\newpage


\subsection*{Introduction: Seeking the most minimal realization of the type-I seesaw model}


The type-I seesaw model~\cite{Minkowski:1977sc,Yanagida:1979as,GellMann:1980vs}
is an elegant and well motivated extension of the Standard Model (SM).
Not only is it capable of explaining the small masses of the SM neutrinos,
it also offers the possibility to account for the observed baryon asymmetry of
the universe via leptogenesis~\cite{Fukugita:1986hr}.
In its usual form, the seesaw mechanism supposes the existence of three right-handed
Majorana neutrinos, $N_I$ ($I=1,2,3$), that participate in Yukawa
interactions with the three charged-lepton flavors, $\ell_\alpha$ ($\alpha = e,\mu,\tau$),
as well as with the SM Higgs, $H$.
This realization of the seesaw mechanism comes with 18 physical parameters
at high energies (nine complex Yukawa couplings $y_{\alpha I}$ plus
three Majorana masses $M_I$ minus three unphysical charged-lepton phases),
which provides enough parametric freedom to reproduce all of the
neutrino oscillation observables measured at low energies,
i.e., the solar and atmospheric mass-squared differences $\Delta m_{\rm sol}^2$
and $\Delta m_{\rm atm}^2$ as well as the three mixing angles $\sin^2\theta_{12}$,
$\sin^2\theta_{13}$, and $\sin^2\theta_{23}$~\cite{Agashe:2014kda,Esteban:2016qun}.


The success of the type-I seesaw model with three right-handed neutrinos 
backs its preeminent role among all conceivable extensions of the Standard Model.
On the other hand, the high dimensionality of the seesaw parameter space
may also be regarded as a drawback, as it diminishes the model's predictive power.
Without further restrictions, the standard seesaw model is, e.g.,
neither able to predict the amount of $CP$ violation in the lepton sector nor the
ordering of the light-neutrino mass spectrum.
This serves as a motivation for studying restricted, more minimal versions
of the seesaw mechanism~\cite{Raidal:2002xf,Guo:2006qa}, in which the size
of the full seesaw parameter space is reduced and which, thus,
boast a larger predictivity.
At any experimental stage, one would, in particular, like to identify the
current most minimal realization of the seesaw model with the
least number of free parameters that is still in accord with the experimental data.
This \textit{most minimal seesaw model} (at a certain experimental stage)
is then singled out by its maximal predictivity, making it an
important benchmark scenario for any future experimental update.%
\footnote{In addition, one may argue that minimal models are more appealing
from the viewpoint of Occam's razor~\cite{Harigaya:2012bw}.}


In this Letter, we are going to explicitly construct such a
\textit{most minimal seesaw model} from a UV perspective.
Our starting point is what is typically referred to as the \textit{minimal seesaw model}:
the ordinary type-I seesaw model featuring only two rather than three
right-handed neutrinos~\cite{Smirnov:1993af,King:1998jw,Frampton:2002qc,Ibarra:2003up}.
Here, note that the case of two right-handed neutrinos is experimentally
still allowed, as we currently lack knowledge of the absolute neutrino mass scale.
It is, hence, still a viable possibility that the lightest SM neutrino
mass eigenstate is in fact massless.
In this case, only two right-handed neutrinos are needed to account
for the nonzero mass-squared differences $\Delta m_{\rm sol}^2$ and $\Delta m_{\rm atm}^2$.
This complies with the fact that successful leptogenesis in the seesaw
model does require two right-handed neutrinos, but not necessarily
more~\cite{Frampton:2002qc,Endoh:2002wm}
(see also \cite{Bjorkeroth:2016qsk,Bambhaniya:2016rbb}).
Within the framework of the minimal seesaw model, we shall now study a particular
UV completion subject to two constraints: (i) a certain Froggatt-Nielsen flavor
symmetry~\cite{Froggatt:1978nt} consistent with the SM charged-lepton mass
hierarchy~\cite{Buchmuller:1998zf} as well as (ii) an exchange symmetry
in the heavy-neutrino sector.
Let us now discuss these two ingredients of our model in turn.


\subsection*{Step 1: Froggatt-Nielsen flavor model for two right-handed neutrinos}


The Yukawa interactions and mass terms of the heavy neutrinos $N_{1,2}$ in the minimal
seesaw model are described by the following Lagrangian (in two-component spinor notation),%
\footnote{We shall work in a basis in which both
the charged-lepton and heavy-neutrino mass matrices are diagonal.}
\begin{align}
\mathcal{L}_{\rm seesaw} = 
- y_{\alpha I}\, \ell_\alpha N_I H - \frac{1}{2}\, M_I N_I N_I + \textrm{h.c.}
\,, \quad \alpha = e,\mu,\tau
\,, \quad I = 1,2 \,.
\label{eq:Lseesaw}
\end{align}
Without further restrictions, this model has 11 physical
parameters at high energies.
One can convince oneself that this is sufficient to fit all of the neutrino
oscillation data, while still keeping some parametric freedom at
low energies~\cite{Branco:2002ie}.
In quest of a \textit{most minimal} realization of the type-I seesaw mechanism,
we would, therefore, like to be more restrictive.
To this end, we shall now embed the Lagrangian in Eq.~\eqref{eq:Lseesaw}
into the Froggatt-Nielsen (FN) model presented in~\cite{Buchmuller:1998zf}.


This flavor model explains the quark and charged-lepton mass hierarchies observed
in the Standard Model as a consequence of a spontaneously broken
Froggatt-Nielsen $U(1)_{\rm FN}$ flavor symmetry. 
In a first step, it supposes that, at energies around the scale of grand unification,
$\Lambda_{\rm GUT} \sim 10^{16}\,\textrm{GeV}$, the SM Yukawa sector is
described by an effective theory with cut-off scale $\Lambda \gtrsim \Lambda_{\rm GUT}$.
In this effective theory, the SM fermions,
$f_i \in \left\{\ell_i,q_i\right\}$,
as well as the SM Higgs are assumed to couple to a SM singlet, the
so-called flavon field $\Phi$, via different (effective) operators,%
\footnote{In realistic models, one typically arrives at such EFTs
after integrating out heavy states with a common mass of $\mathcal{O}\left(\Lambda\right)$
that couple to both: SM fermion-Higgs pairs as well as to the flavon;
see \cite{Babu:2009fd} and references therein.}
\begin{align}
\mathcal{L}_{\rm eff} \supset a_{ij}
\left(\frac{\Phi}{\Lambda}\right)^{p_{ij}} f_i f_j\, H \,.
\label{eq:LeffFN}
\end{align}
Here, $i$ is a generic flavor index and the $a_{ij}$ are dimensionless,
undetermined coefficients of $\mathcal{O}\left(1\right)$.
The fermions and the flavon $\Phi$ are charged under an Abelian $U(1)_{\rm FN}$ flavor
symmetry, with the flavon carrying in particular a charge of $\left[\Phi\right]_{\rm FN} = -1$.
Insisting on $U(1)_{\rm FN}$ being a good (global or local) symmetry of the effective theory,
the FN charge assignment to the fermions fixes the powers $p_{ij}$ in
Eq.~\eqref{eq:LeffFN}, $p_{ij} = \left[f_i\right]_{\rm FN} + \left[f_j\right]_{\rm FN}$.
In a second step, $\Phi$ is assumed to acquire a nonzero
vacuum expectation value, $\left<\Phi\right>$, at energies slightly below $\Lambda$,
thereby breaking $U(1)_{\rm FN}$ spontaneously.
This generates the SM Yukawa couplings.
At this point, the ratio $\left<\Phi\right> / \Lambda$ defines a universal small parameter,
which provides a useful parametrization of the SM Yukawa matrices,
\begin{align}
\Phi \rightarrow \left<\Phi\right> \quad\Rightarrow\quad 
\mathcal{L}_{\rm eff} \rightarrow y_{ij} f_i f_j\, H \,, \quad
y_{ij} = a_{ij}\, \epsilon_0^{p_{ij}} \,, \quad \epsilon_0 = \frac{\left<\Phi\right>}{\Lambda} \,.
\end{align}
The authors of~\cite{Buchmuller:1998zf} focus on FN charges that commute with $SU(5)$.
They group the SM fermions into
representations of $SU(5)$ and assign universal charges to each multiplet,
see Tab.~\ref{tab:FNcharges}.
Setting the FN hierarchy parameter $\epsilon_0$ to a particular value,
$\epsilon_0 \simeq 0.17$, this simple model then manages to reproduce
the phenomenology of the SM quark and charged-lepton mass matrices.


\begin{table}
\begin{center}
\begin{tabular}{|c||cccccccc|c|} \hline
Field  & $\mathbf{10}_1$  & $\mathbf{10}_2$  & $\mathbf{10}_3$  &
         $\mathbf{5}_1^*$ & $\mathbf{5}_2^*$ & $\mathbf{5}_3^*$ & 
         $\mathbf{1}_1$   & $\mathbf{1}_2$   & $\Phi$              \\\hline\hline
Charge & $2$ & $1$ & $0$ & $1$ & $0$ & $0$ & $q$ & $q$ & $-1$      \\\hline
\end{tabular}
\caption{Assignment of Froggatt-Nielsen flavor charges to the SM fermions, right-handed
neutrinos as well as to the flavon field $\Phi$~\cite{Buchmuller:1998zf}.
Here, the SM fermions and right-handed neutrinos are grouped into $SU(5)$ representations:
$\mathbf{10} = \left(q,u,e\right)$, $\mathbf{5}^* = \left(d,\ell\right)$,
$\mathbf{1}_1 = N_1 $ and $\mathbf{1}_2 = N_2$.
The right-handed neutrino charge $q$ is undetermined at first.}
\label{tab:FNcharges}
\end{center}
\end{table}


As indicated in Tab.~\ref{tab:FNcharges}, we shall now extend the FN
model of~\cite{Buchmuller:1998zf} to the heavy-neutrino sector.
In doing so, let us assume that the FN dynamics are not only responsible
for the structure of the heavy-neutrino Yukawa couplings, but
also for the hierarchy among the heavy-neutrino masses, 
\begin{align}
y_{\alpha I} = a_{\alpha I}\, \epsilon_0^{p_{\alpha I}} \,, \quad 
p_{\alpha I} = \left[\ell_\alpha\right]_{\rm FN} + \left[N_I\right]_{\rm FN} \,, \quad
M_I = b_{I}\,\epsilon_0^{q_I} M_0 \,, \quad q_I = 2\left[N_I\right]_{\rm FN} \,.
\label{eq:yM}
\end{align}
Here, $a_{\alpha I}$ and $b_I$ are again dimensionless coefficients of $\mathcal{O}\left(1\right)$,
while $M_0$ is a large mass scale, which may, e.g., be generated
in the course of spontaneous $U(1)_{B-L}$ breaking.
In view of the expression for $M_I$ in Eq.~\eqref{eq:yM}, we decide 
to assign equal charges to the two right-handed neutrinos,
$\left[N_1\right]_{\rm FN} = \left[N_2\right]_{\rm FN} = q\geq 0$.
This is motivated by the fact that successful leptogenesis in
the minimal seesaw model requires an approximate degeneracy between the two
Majorana masses, $M_1 \simeq M_2$~\cite{Bjorkeroth:2016qsk,Bambhaniya:2016rbb}.
With the complete FN charge assignment given in Tab.~\ref{tab:FNcharges},
we now expect the neutrino Yukawa matrix $y_{\alpha I}$ in Eq.~\eqref{eq:Lseesaw}
to exhibit the following hierarchy structure,
\begin{align}
y_{\alpha I} \sim
\begin{pmatrix}
\epsilon_0 & \epsilon_0 \\
1 & 1 \\
1 & 1
\end{pmatrix} y_0 \,, \quad y_0 \equiv \epsilon_0^q \,,
\label{eq:yepsilon}
\end{align}
where each entry comes with an $\mathcal{O}\left(1\right)$ uncertainty
and where $y_0$ quantifies the universal suppression of all Yukawa
couplings because of the right-handed neutrino charge $q$.
To be more precise, we may explicitly parametrize the Yukawa matrix
$y_{\alpha I}$ in our model as follows,
\begin{align}
y_{\alpha I} =
\begin{pmatrix}
\:\:\,\epsilon\,\cos\theta_e\, e^{i\varphi_e} &
\:\:\,\epsilon\,\sin\theta_e\,e^{i\left(\varphi_e+\Delta\varphi_e\right)} \\
\phantom{c_{\mu\tau}}\cos\theta_\mu\,e^{i\varphi_\mu} &
\phantom{c_{\mu\tau}}\sin\theta_\mu\,e^{i\left(\varphi_\mu+\Delta\varphi_\mu\right)} \\
c_{\mu\tau}\cos\theta_\tau\, e^{i\varphi_\tau} &
c_{\mu\tau}\sin\theta_\tau\,e^{i\left(\varphi_\tau+\Delta\varphi_\tau\right)} \\
\end{pmatrix} y_0 \,.
\label{eq:yalphaI}
\end{align}
Here, $-\frac{\pi}{2} \leq \theta_\alpha \leq \frac{\pi}{2}$ denote three ``mixing angles'',
$0 \leq \varphi_\alpha \leq 2\pi$ and $0 \leq \Delta\varphi_\alpha \leq 2\pi$ are
three phases and phase shifts, respectively, and $c_{\mu\tau}$ is a dimensionless
number of $\mathcal{O}\left(1\right)$.
The FN hierarchy parameter $\epsilon$ is expected to take a value close
to the one deduced in~\cite{Buchmuller:1998zf}, $\epsilon \simeq \epsilon_0 \simeq 0.17$.
But in order to remain conservative, we shall also allow for small
deviations from this value.


\subsection*{Step 2: Exchange symmetry in the heavy-neutrino Yukawa and mass terms}


\begin{table}
\begin{center}
\begin{tabular}{|c||cccc|}\hline
Observable & Units & Hierarchy & Best-fit value & $3\,\sigma$ confidence interval \\\hline\hline
$\Delta m_{21}^2$ & $\left[10^{-5}\,\textrm{eV}^2\right]$
& \vphantom{\bigg(} both & $+7.50$ & $\left[+7.03,+8.09\right]$ \\\hline
\multirow{2}{*}{$\Delta m_{3\ell}^2$} &
\multirow{2}{*}{$\left[10^{-3}\,\textrm{eV}^2\right]$}
& \vphantom{\Big(} NH & $+2.52$ & $\left[+2.41,+2.64\right]$ \\
& & \vphantom{\Big(} IH & $-2.51$ & $\left[-2.64, -2.40\right]$ \\\hline
$\sin^2\theta_{12}$ & $\left[10^{-1}\right]$
& \vphantom{\bigg(} both & $+3.06$ & $\left[+2.71,+3.45\right]$ \\\hline
\multirow{2}{*}{$\sin^2\theta_{13}$} &
\multirow{2}{*}{$\left[10^{-2}\right]$}
& \vphantom{\Big(} NH & $+2.17$ & $\left[+1.93,+2.39\right]$ \\
& & \vphantom{\Big(} IH & $+2.18$ & $\left[+1.95,+2.41\right]$ \\\hline
\multirow{2}{*}{$\sin^2\theta_{23}$} &
\multirow{2}{*}{$\left[10^{-1}\right]$}
& \vphantom{\Big(} NH & $+4.41$ & $\left[+3.85,+6.35\right]$  \\
& & \vphantom{\Big(} IH & $+5.87$ & $\left[+3.93,+6.40\right]$ \\\hline
\multirow{2}{*}{$\delta$} &
\multirow{2}{*}{$\left[\textrm{deg}\right]$}
& \vphantom{\Big(} NH & $261$ & $\left[0,360\right]$ \\
& & \vphantom{\Big(} IH & $277$ & $\left[0,31\right]\oplus\left[145,360\right]$
\\\hline
\end{tabular}
\caption{Best-fit values and $3\,\sigma$ confidence intervals for the five low-energy
observables that are currently accessible in experiments
($\Delta m_{21}^2$, $\Delta m_{3\ell}^2$, $\sin^2\theta_{12}$,
$\sin^2\theta_{13}$, and $\sin^2\theta_{23}$) as well as for the $CP$-violating phase
$\delta$~\cite{Esteban:2016qun}.
Here, we define $\Delta m_{ij}^2 \equiv m_i^2 - m_j^2$ in general and
$\Delta m_{3\ell}^2 \equiv \Delta m_{31}^2 > 0$ for NH and
$\Delta m_{3\ell}^2 \equiv \Delta m_{32}^2 < 0$ for IH in particular.
}
\label{tab:data}
\end{center}
\end{table}


Our expression for $y_{\alpha I}$ in Eq.~\eqref{eq:yalphaI} trivially exhibits 
the same number of undetermined parameters as the most general $3\times 2$ Yukawa matrix:
six absolute values ($y_0$, $\epsilon$, $c_{\mu\tau}$, and $\theta_\alpha$) plus six phases
($\varphi_\alpha$ and $\Delta\varphi_\alpha$).
More work is, therefore, needed to arrive at a more minimal
realization of the type-I seesaw mechanism.
Inspired by the FN charge assignment in Tab.~\ref{tab:data}
as well as by the resulting estimate for $y_{\alpha I}$ in Eq.~\eqref{eq:yepsilon},
we are now going to make our second model assumption:
We suppose an approximate exchange symmetry in the heavy-neutrino
Yukawa (and mass) terms,
\begin{align}
\left|y_{\alpha 1}\right| \approx \left|y_{\alpha 2}\right| \,, \quad
M_1 \approx M_2 \,,
\label{eq:exchange}
\end{align}
which means that $\tan\theta_\alpha \approx 1$ for all flavors.
Up to small perturbations, this reduces the number of independent
absolute values in $y_{\alpha I}$ from six ($y_0$, $\epsilon$, $c_{\mu\tau}$,
and $\theta_\alpha$) to three ($y_0$, $\epsilon$, and $c_{\mu\tau}$).


The assumption in Eq.~\eqref{eq:exchange} immediately entails the question
as to whether such a Yukawa matrix is still capable of accounting for all of the
low-energy observables.
For an \textit{exact} exchange symmetry the answer is negative,
as it would always be inconsistent with three nonzero mixing angles.
An approximate exchange symmetry,
$\left|y_{\alpha 1}\right| = \left|y_{\alpha 2}+\delta y_\alpha\right|$,
is, by contrast, viable---simply because in this case, the small
perturbations around the symmetric ``leading-order'' Yukawa couplings,
$\left|\delta y_\alpha\right| \ll \left|y_{\alpha I}\right|$,
have their share in reproducing the low-energy oscillation data.
To see this explicitly, it is best to employ the
Casas-Ibarra parametrization (CIP)~\cite{Casas:2001sr}
for the neutrino Yukawa matrix $y_{\alpha I}$.
In the case of only two right-handed neutrinos, the CIP
can be brought into the following compact and
dimensionless form (see~\cite{Rink:2016vvl} for more details),
\begin{align}
\kappa_{\alpha 1} = \frac{1}{\sqrt{2}}
\left(V_\alpha^+\, e^{-iz} + V_\alpha^-\, e^{+iz}\right) \,, \quad 
\kappa_{\alpha 2} = \frac{i}{\sqrt{2}}
\left(V_\alpha^-\, e^{+iz} - V_\alpha^+\, e^{-iz}\right) \,,
\label{eq:CI}
\end{align}
where $z$ is an arbitrary complex number and where $\kappa_{\alpha I}$
and $V_\alpha^\pm$ are defined as follows,
\begin{align}
\kappa_{\alpha I} = \sqrt{\frac{v_{\rm ew}}{M_I}} \: y_{\alpha I} \,, \quad
V_{\alpha}^\pm = \frac{1}{\sqrt{2}}\left(V_{\alpha k} \pm i\, V_{\alpha l}\right) \,, \quad
V_{\alpha i} = \sqrt{\frac{m_i}{v_{\rm ew}}}\, i\, U_{\alpha i}^* \,.
\label{eq:defs}
\end{align}
Here, $v_{\rm ew} \simeq 174\,\textrm{GeV}$ denotes the electroweak scale,
$m_i$ are the SM neutrino mass eigenvalues, and $U$ is the
Pontecorvo-Maki-Nakagawa-Sakata (PMNS) lepton mixing
matrix~\cite{Pontecorvo:1957qd,Maki:1962mu}.
The neutrino indices $\left(k,l\right)$ need to be chosen as $\left(2,3\right)$
in the case of a normal mass hierarchy (NH), $m_1 < m_2 < m_3$, and as $\left(1,2\right)$
in the case of an inverted mass hierarchy (IH), $m_3 < m_1 < m_2$.
With these definitions, one recognizes $\kappa_{\alpha I}$ as dimensionless
and rescaled Yukawa couplings, while the matrix $V$ plays the role of a
rescaled version of the PMNS matrix.
The advantage of this version of the CIP
is that it separates the unknown model parameters at high energies
(left-hand side of Eq.~\eqref{eq:CI}) from the observables that one
can measure at low energies (right-hand side of Eq.~\eqref{eq:CI}).
This means, in particular, that one can study the rescaled Yukawa
couplings $\kappa_{\alpha I}$ as functions of the low-energy observables
without knowledge of the heavy-neutrino masses $M_{1,2}$.


\begin{figure}
\begin{center}
\includegraphics[width=0.31\textwidth]{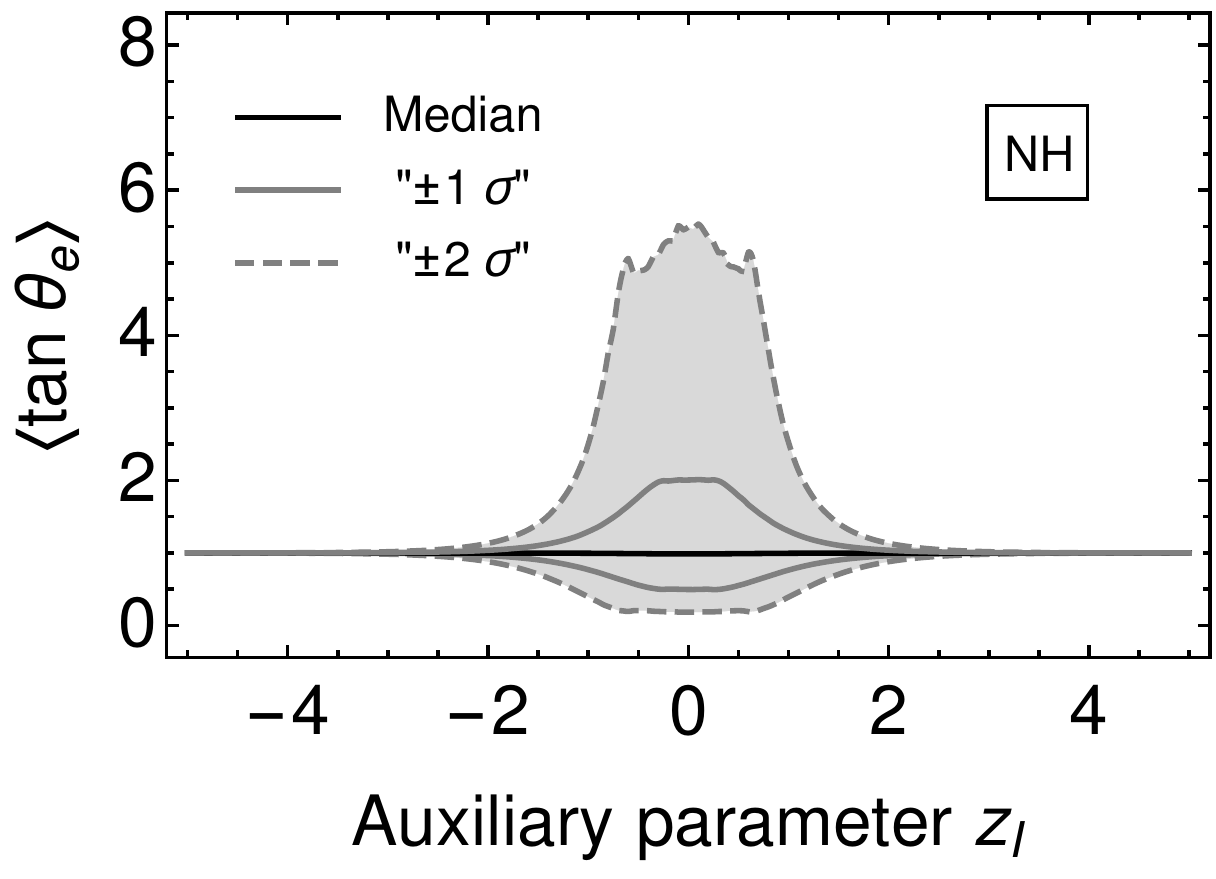}  \hspace{0.01\textwidth}
\includegraphics[width=0.31\textwidth]{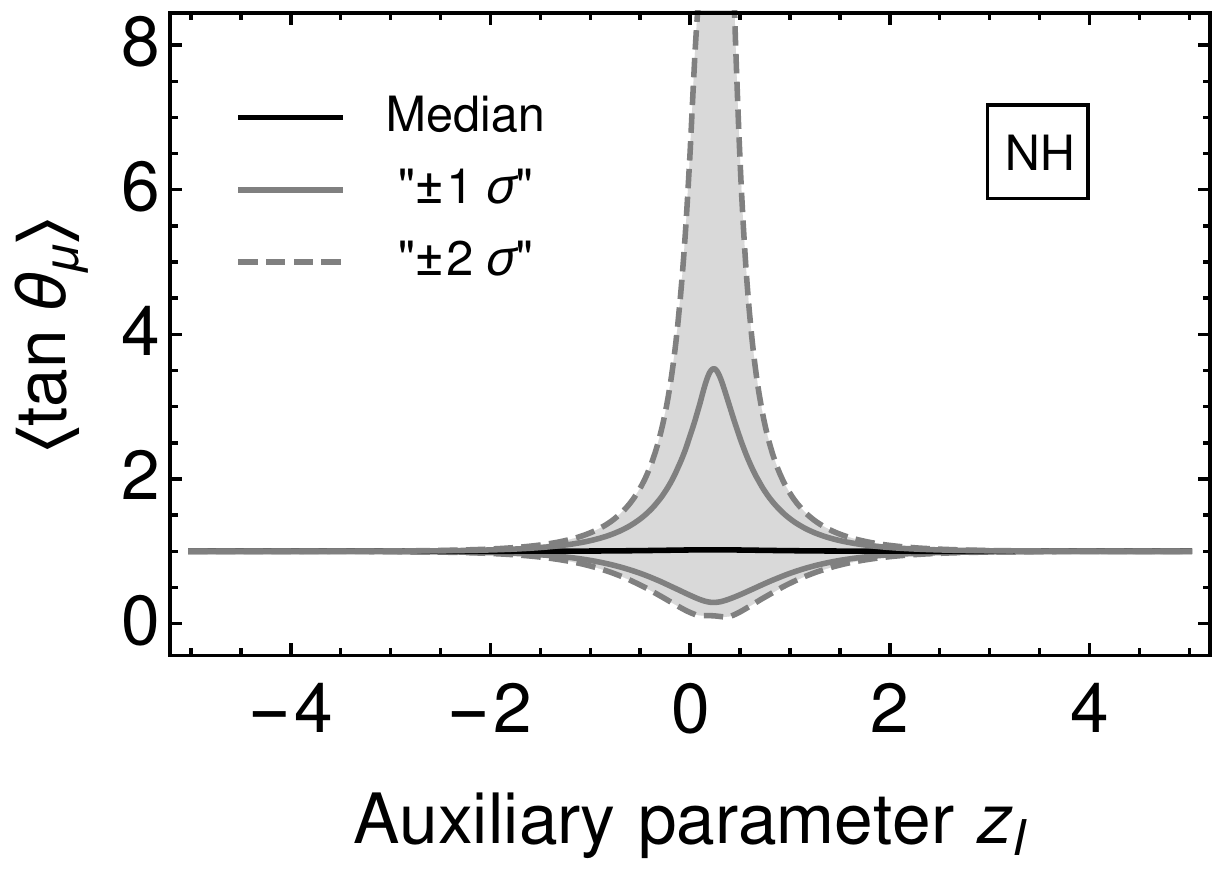} \hspace{0.01\textwidth}
\includegraphics[width=0.31\textwidth]{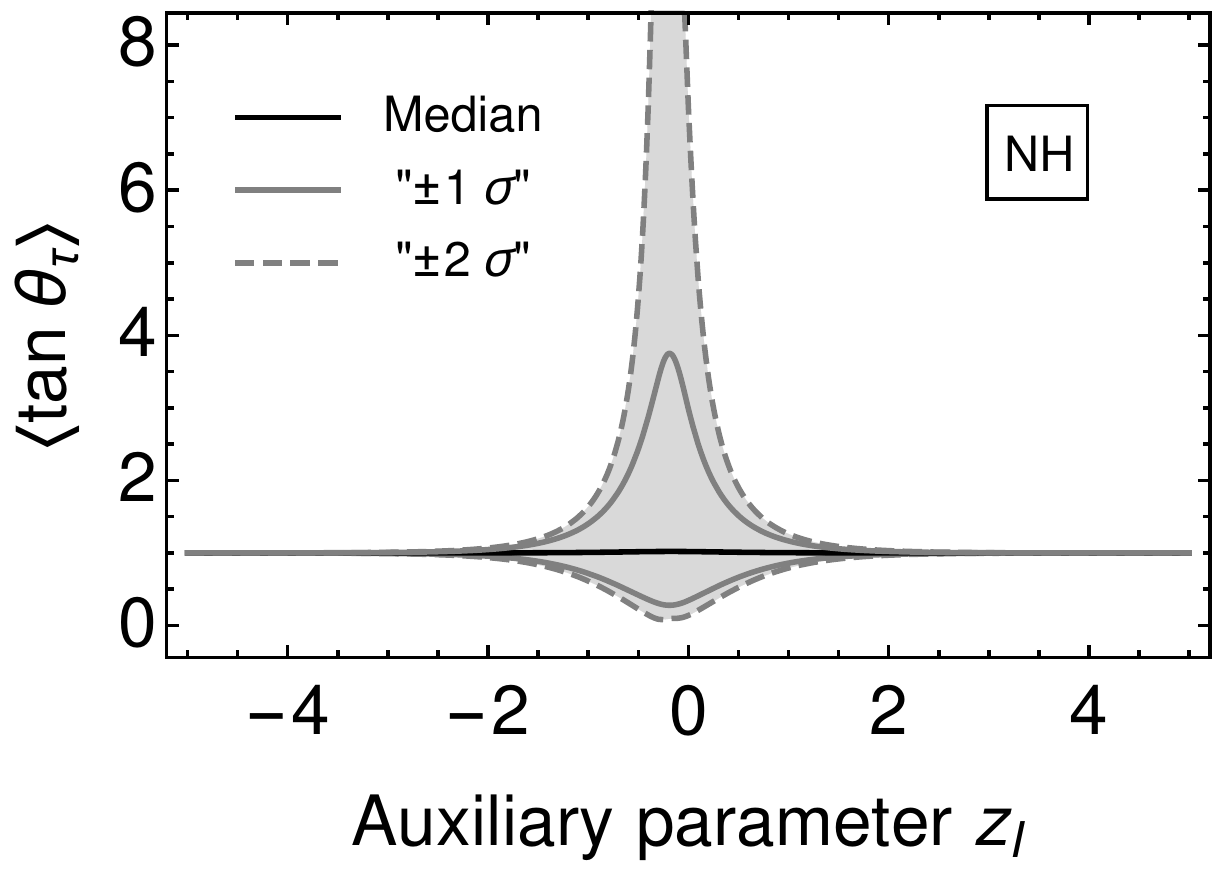}

\bigskip
\includegraphics[width=0.31\textwidth]{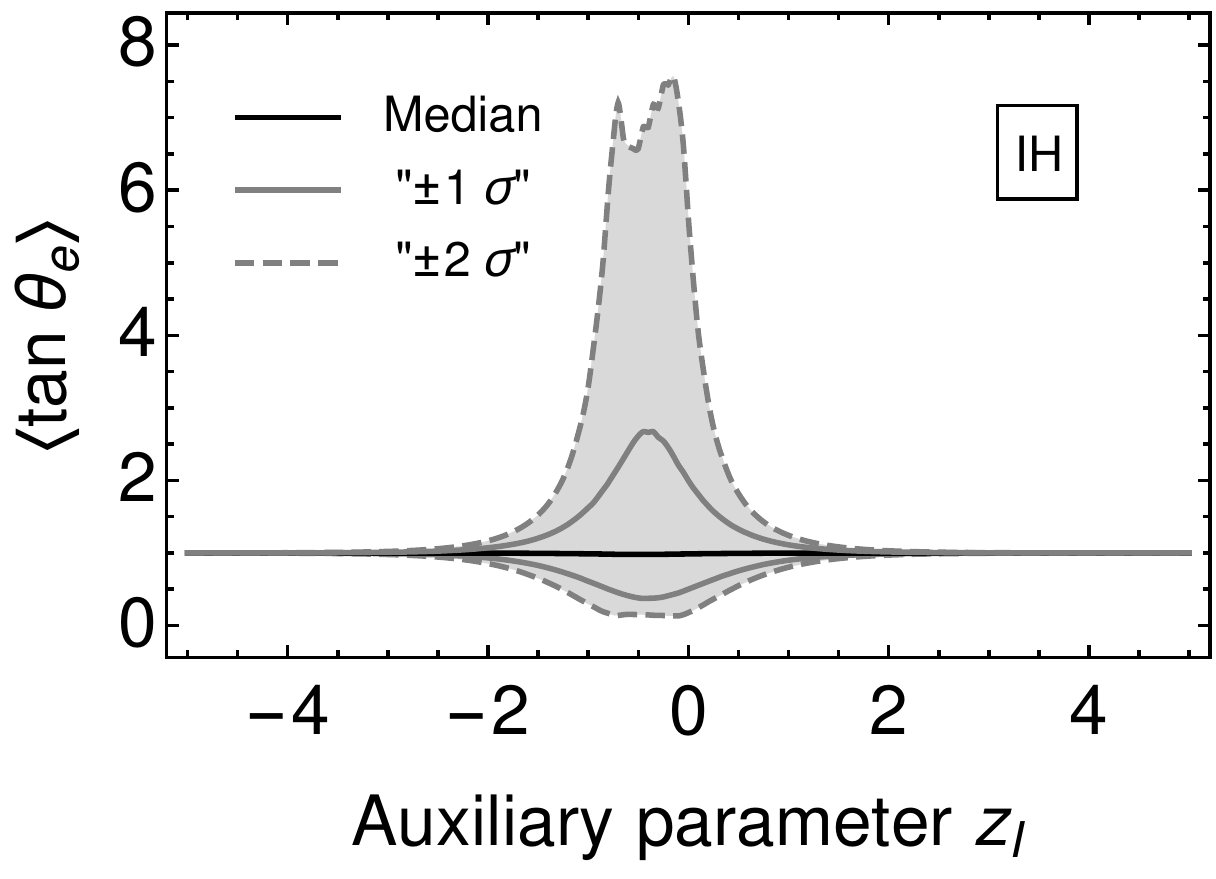}  \hspace{0.01\textwidth}
\includegraphics[width=0.31\textwidth]{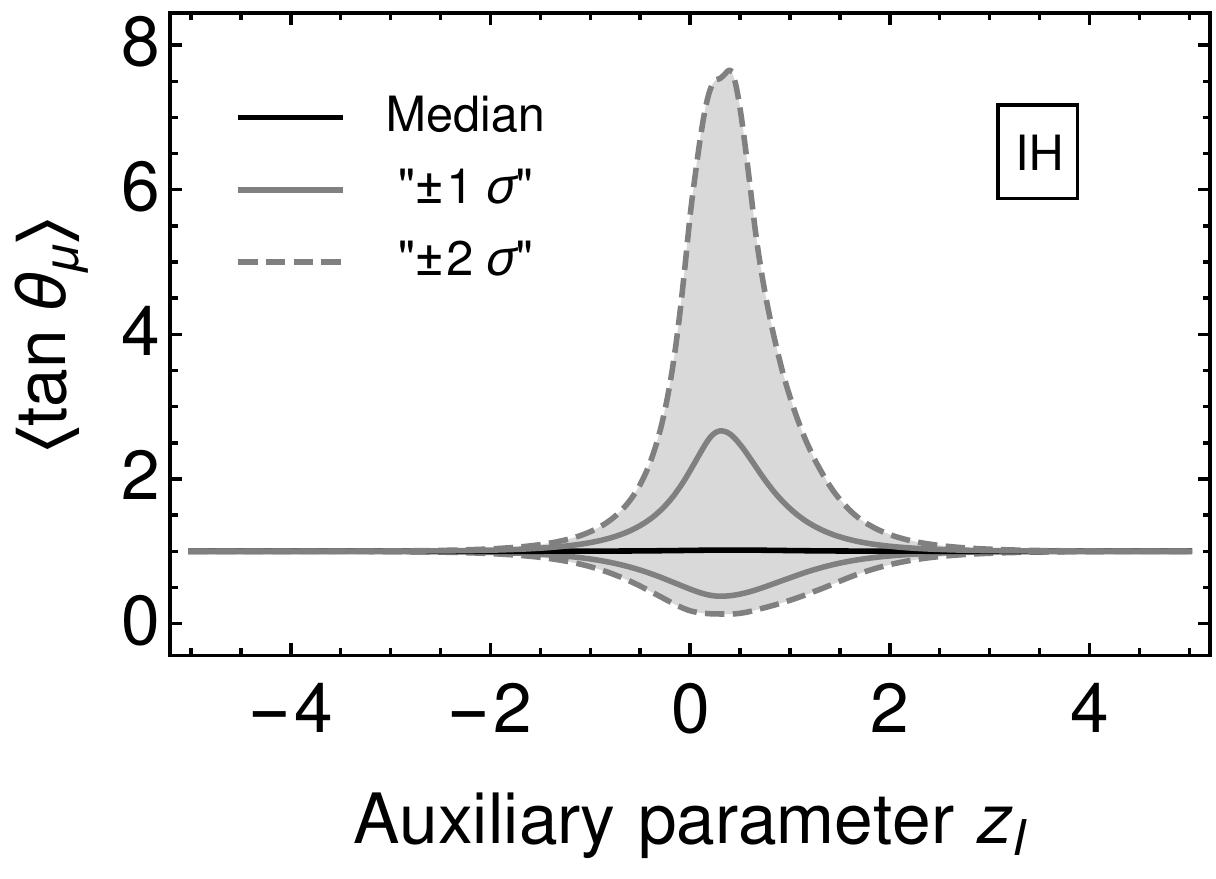} \hspace{0.01\textwidth}
\includegraphics[width=0.31\textwidth]{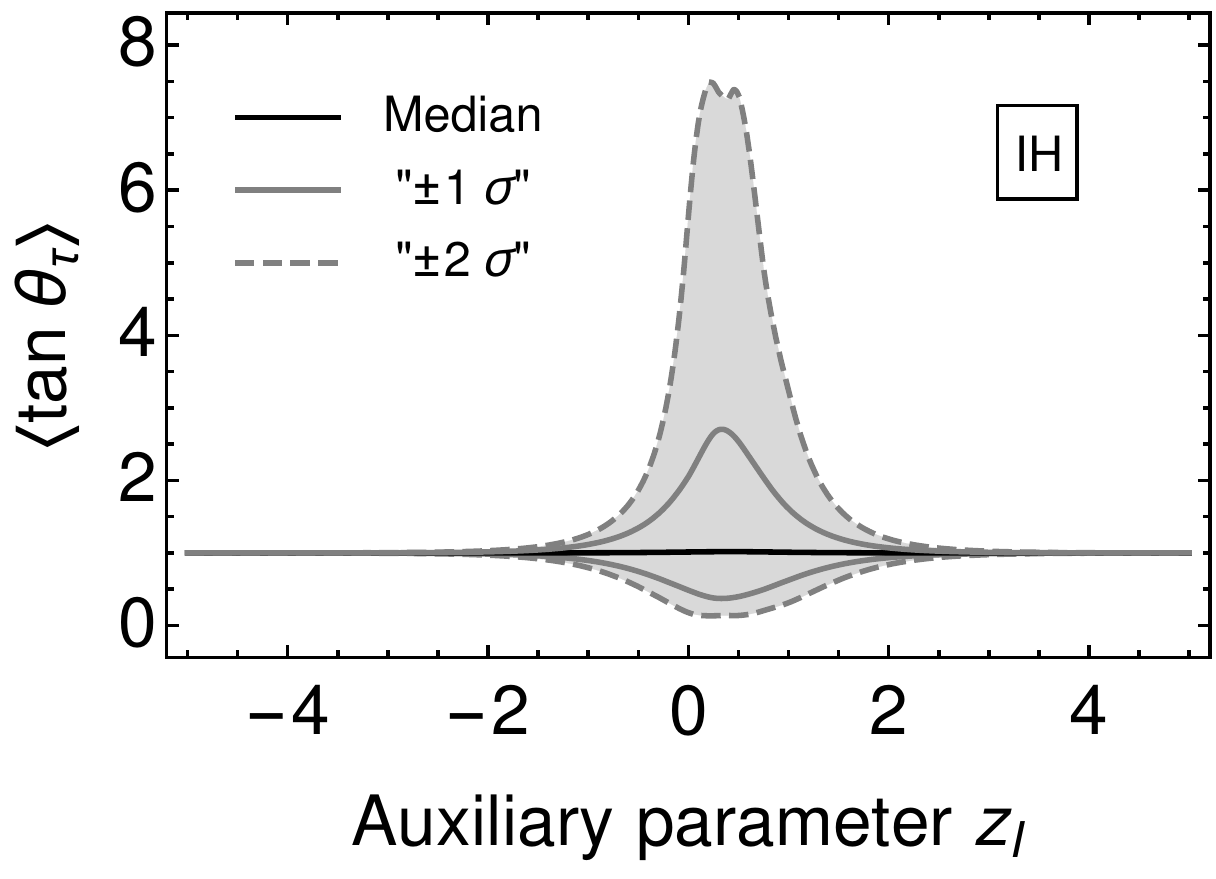}
\caption{Expectation values of $\tan\theta_e$, $\tan\theta_\mu$, and $\tan\theta_\tau$
as functions of the imaginary component of the auxiliary parameter $z$ in the CIP
for both NH \textbf{(upper panel)} and IH \textbf{(lower panel)}.
See Eq.~\eqref{eq:yalphaI} for the definition of the mixing angles $\theta_\alpha$ and
Eq.~\eqref{eq:CI} for the definition of $z$.
For each value of $z_I$, we generate distributions of possible $\tan\theta_\alpha$ values,
varying the auxiliary parameter $z_R$ as well as the two $CP$-violating phases in the
lepton mixing matrix, $\delta$ and $\sigma$, over their entire ranges of allowed values.
All other observables are fixed at their respective best-fit values, see Tab.~\ref{tab:data}.
The curves in the above plots indicate the following quantiles $Q_p$ of our $\tan \theta_\alpha$
distributions:\newline
\{$-2\,\sigma$, $-1\,\sigma$, median, $+1\,\sigma$, $+2\,\sigma$\}
$\rightarrow p = \left\{2.28,15.87,50.00,84.13,97.72\right\}\,\%$, in analogy to
a Gaussian distribution.}
\label{fig:tan} 
\end{center}
\end{figure}


From Eq.~\eqref{eq:CI}, it is now immediately evident that the parameter
space of the minimal seesaw model does, indeed, encompass regions in
which the requirement in Eq.~\eqref{eq:exchange} is satisfied.
All we have to do is to give the auxiliary parameter
$z = \left(z_R + i\, z_I\right)/\sqrt{2}$ a large imaginary part $z_I$,
\begin{align}
z_I \gg 1 \:\:\Rightarrow\:\:
\kappa_{\alpha 1} \simeq i\,\kappa_{\alpha 2}
\simeq \frac{1}{\sqrt{2}}\, V_\alpha^+\,e^{-iz} \,, \qquad
z_I \ll 1 \:\:\Rightarrow\:\:
\kappa_{\alpha 1} \simeq -i\,\kappa_{\alpha 2}
\simeq \frac{1}{\sqrt{2}}\, V_\alpha^-\,e^{+iz} \,.
\label{eq:alignment}
\end{align}
That is to say that, for $\left|z_I\right| \gg 1$, we obtain
$\left|\kappa_{\alpha 1}\right| \simeq \left|\kappa_{\alpha 2}\right|$.
Thanks to the approximate degeneracy among the heavy-neutrino masses in our model,
$M_1 \simeq M_2$, this readily implies
$\left|y_{\alpha 1}\right| \simeq \left|y_{\alpha 2}\right|$.
In~\cite{Rink:2016vvl}, this situation is referred to as
\textit{flavor alignment}, which reflects the fact that,
for similar column vectors in $y_{\alpha I}$, the linear
combinations of charged-lepton flavors coupling to $N_1$ and $N_2$,
respectively, $\ell_I \propto y_{\alpha I}\, \ell_\alpha$,
are closely aligned to each other in the $\left(e,\mu,\tau\right)$ flavor space.
To demonstrate more explicitly how large values of $\left|z_I\right|$ lead to flavor
alignment in the neutrino Yukawa matrix, we may also study the $z_I$ dependence
of the angles $\theta_\alpha$ in Eq.~\eqref{eq:yalphaI}.
The outcome of this analysis is shown in Fig.~\ref{fig:tan}, which confirms that,
for large enough $\left|z_I\right|$, the condition in Eq.~\eqref{eq:exchange}
is more or less fulfilled.
For $\left|z_I\right| \gtrsim 2$, all $\tan\theta_\alpha$ expectation
values converge to unity.


A further lesson from Eq.~\eqref{eq:alignment} is that flavor alignment
in the neutrino Yukawa matrix can only be realized at the cost
of a specific phase relation between $y_{\alpha 1}$ and $y_{\alpha 2}$,
\begin{align}
y_{\alpha 1} \simeq s\,i\,y_{\alpha 2} \,,
\quad s = \textrm{sign}\left\{z_I\right\} \,,
\end{align}
which leaves us (up to a discrete sign) with only one choice for the
phase shifts in Eq.~\eqref{eq:yalphaI},  $\Delta\varphi_\alpha \simeq -s\,\pi/2$.
Modulo small perturbations, the assumption of an approximate exchange symmetry
in the Yukawa matrix, therefore, eliminates six parameters: three angles,
$\theta_\alpha$, as well as three phase shifts, $\Delta\varphi_\alpha$.
On top of that, one can always absorb the remaining three phases, $\varphi_\alpha$,
into the charged-lepton fields.
In the limit of an exact exchange symmetry, the number of free parameters
in the neutrino Yukawa matrix, thus, reduces to three ($y_0$, $\epsilon$, and $c_{\mu\tau}$).
Out of these, $y_0$ and $\epsilon$ can, however, be estimated
within the context of  our flavor model.
Therefore, with the $\mathcal{O}\left(1\right)$ parameter $c_{\mu\tau}$ remaining
as the only parameter in the neutrino Yukawa matrix that we do \textit{not} have a
theoretical handle on, this truly is a most minimal realization
of the type-I seesaw model!
In summary, we conclude that, in consequence of our two model
assumptions, the seesaw Lagrangian in
Eq.~\eqref{eq:Lseesaw} can now approximately be written as follows,
\begin{align}
\mathcal{L}_{\rm seesaw} \sim
- y_0 \left(\epsilon\,\ell_e + \ell_\mu + c_{\mu\tau}\, \ell_\tau\right)
\left(N_1 - s\,i\,N_2\right) H
- \frac{1}{2}\,M \left(N_1 N_1 + N_2 N_2 \right) + \textrm{h.c.} \,,
\label{eq:Lmodel}
\end{align}
where $M = \left(M_1+M_2\right)/2 \simeq M_1 \simeq M_2$.
This seesaw Lagrangian is minimal for three reasons:
\begin{enumerate}
\item It features only two rather than three right-handed neutrinos.
Moreover, as required by successful leptogenesis, the masses of these neutrinos
are approximately degenerate, meaning that $N_1$ and $N_2$
form a pair of pseudo-Dirac neutrinos (with Dirac mass $M$).
\item The Yukawa matrix $y_{\alpha I}$ exhibits flavor alignment,
i.e., modulo small perturbations, it contains only three free parameters
($y_0$, $\epsilon$, and $c_{\mu\tau}$).
The heavy-neutrino Yukawa interactions are, thus, invariant
under the approximate exchange symmetry $N_1 \leftrightarrow - s\,i\,N_2$.
This needs to be contrasted with the heavy-neutrino mass terms in Eq.~\eqref{eq:Lmodel},
which do \textit{not} obey this symmetry.
The requirement of nearly degenerate Majorana masses, $M_1 \simeq M_2$,
implies that the mass sector needs to respect a slightly
different exchange symmetry, $N_1 \leftrightarrow N_2$.
\item As a consequence of our Froggatt-Nielsen model, the $N_{1,2}$ coupling to
the electron flavor is parametrically suppressed compared to the muon and tau
flavors, $\epsilon \ll 1$.
In other words, the neutrino Yukawa matrix exhibits an approximate
two-zero texture (see also~\cite{Rink:2016vvl}).
\end{enumerate}
In our discussion up to this point, we argued that the Lagrangian in
Eq.~\eqref{eq:Lmodel} follows from two physical
assumptions: (i) a Froggatt-Nielsen flavor symmetry with charges as
listed in Tab.~\ref{tab:FNcharges}
as well as (ii) an approximate exchange symmetry in the heavy-neutrino Yukawa
and mass terms.
At the same time, we demonstrated that Eq.~\eqref{eq:Lmodel} 
can also be obtained in the large-$\left|z_I\right|$ limit of
the CIP. 
This indicates that, despite its extreme minimality, the Lagrangian
$\mathcal{L}_{\rm seesaw}$ in Eq.~\eqref{eq:Lmodel} ought to be able
to reproduce all of the low-energy neutrino data.
We will now show that this is, indeed, the case and discuss further predictions
that one can deduce from Eq.~\eqref{eq:Lmodel}.


\subsection*{Predictions: Normal light-neutrino
mass hierarchy and maximal \boldmath{$CP$ violation}}


\begin{figure}
\begin{center}
\includegraphics[width=0.46\textwidth]{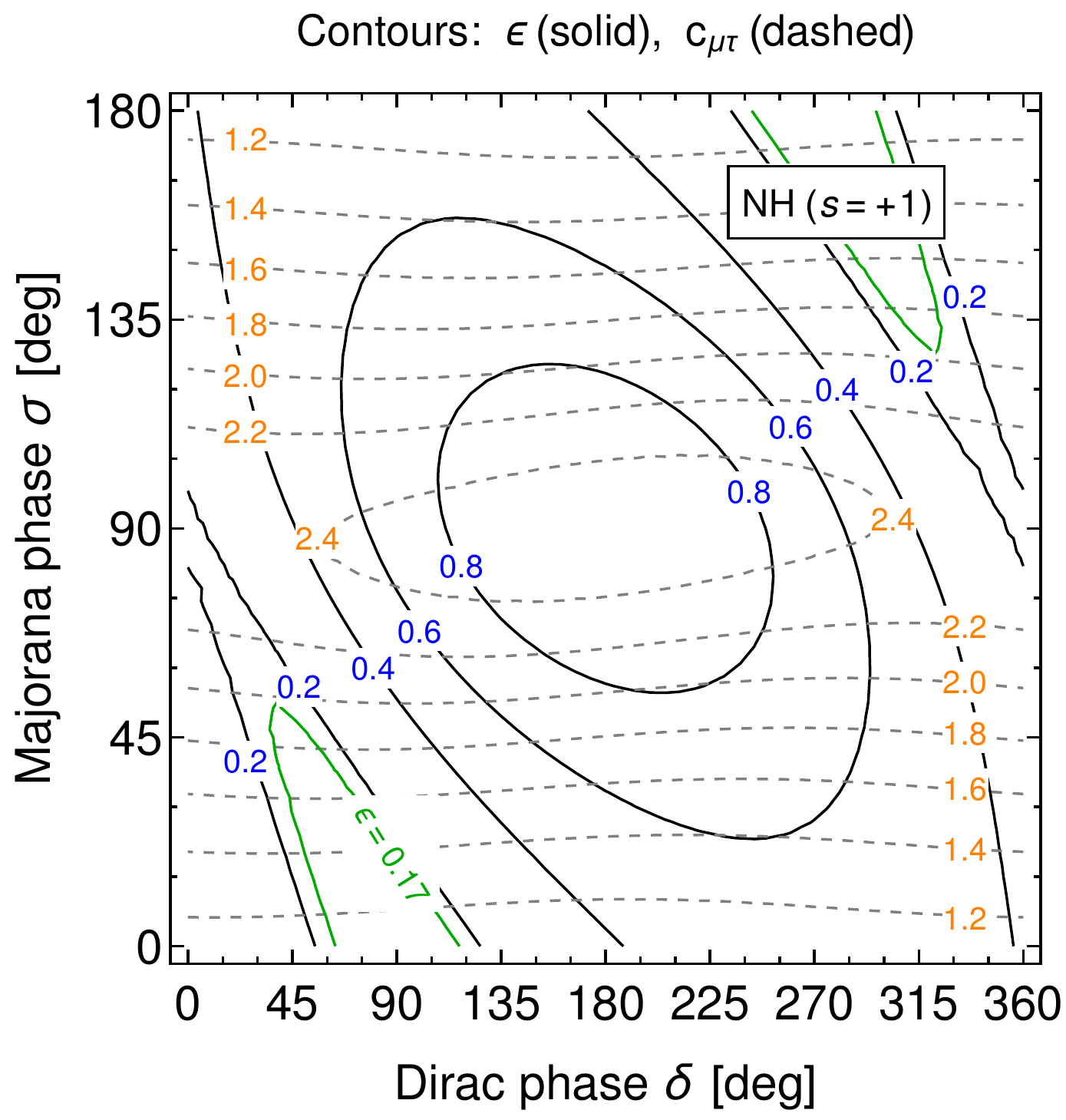}  \hspace{0.05\textwidth}
\includegraphics[width=0.46\textwidth]{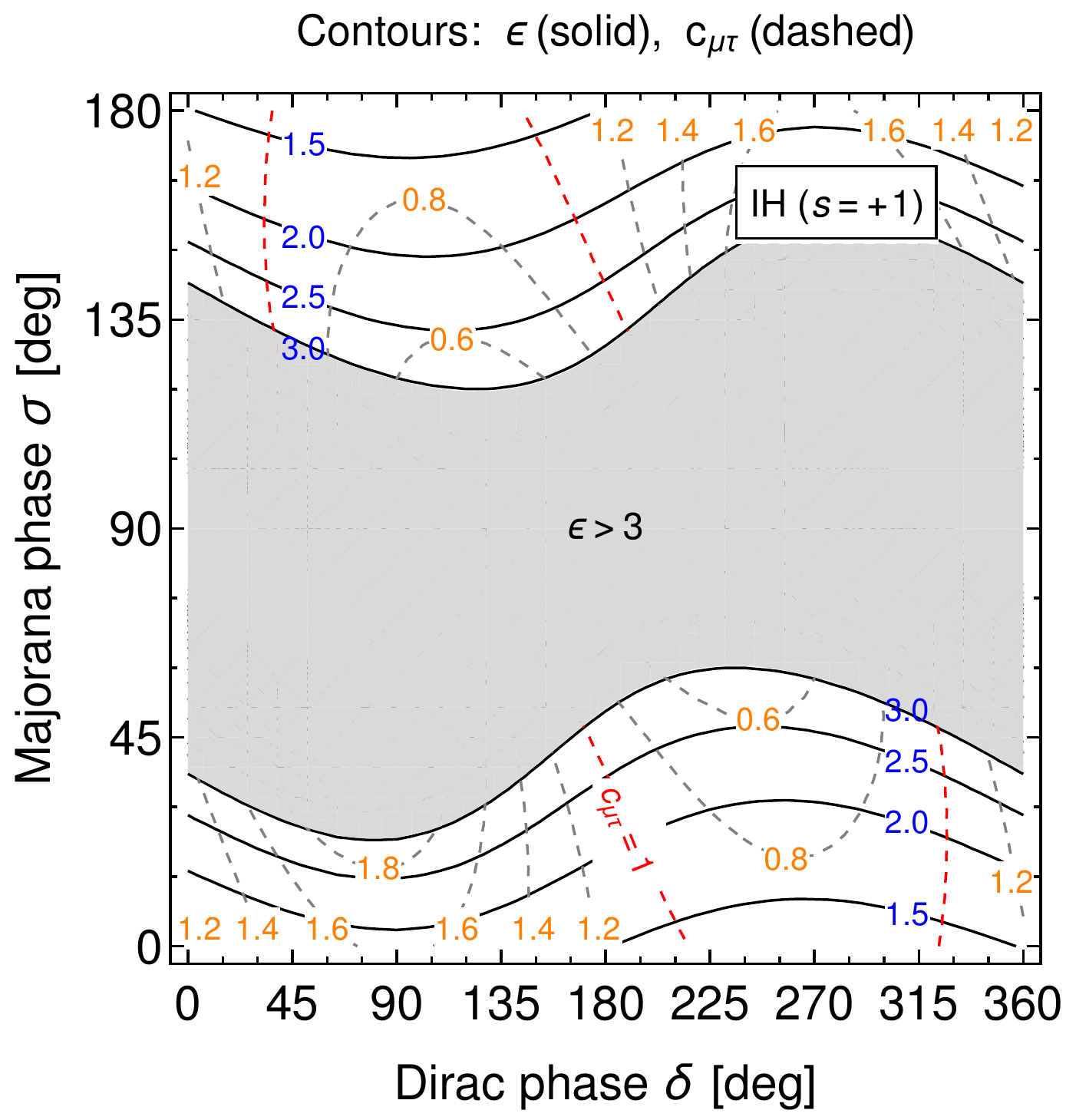}

\bigskip
\includegraphics[width=0.46\textwidth]{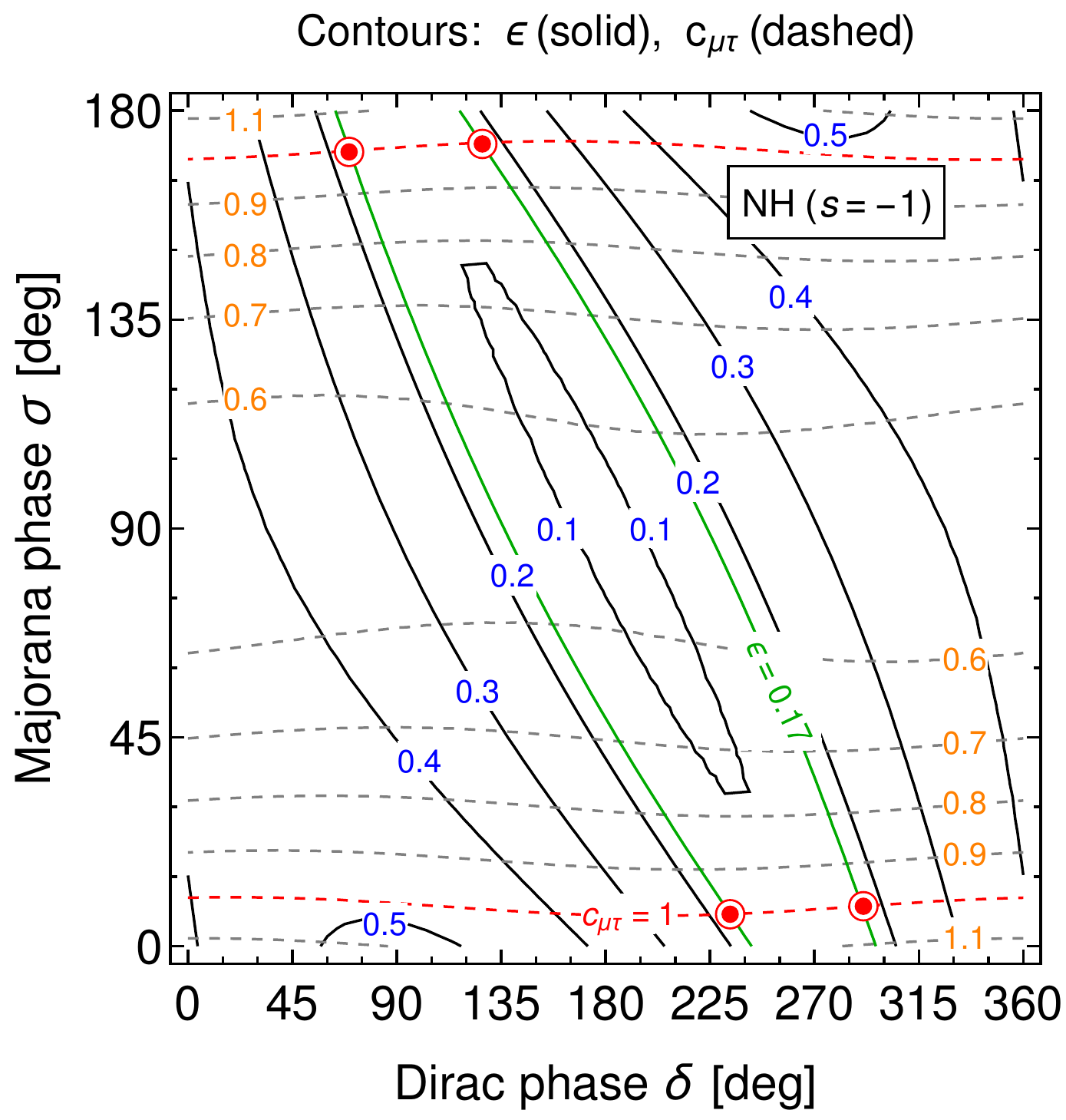}  \hspace{0.05\textwidth}
\includegraphics[width=0.46\textwidth]{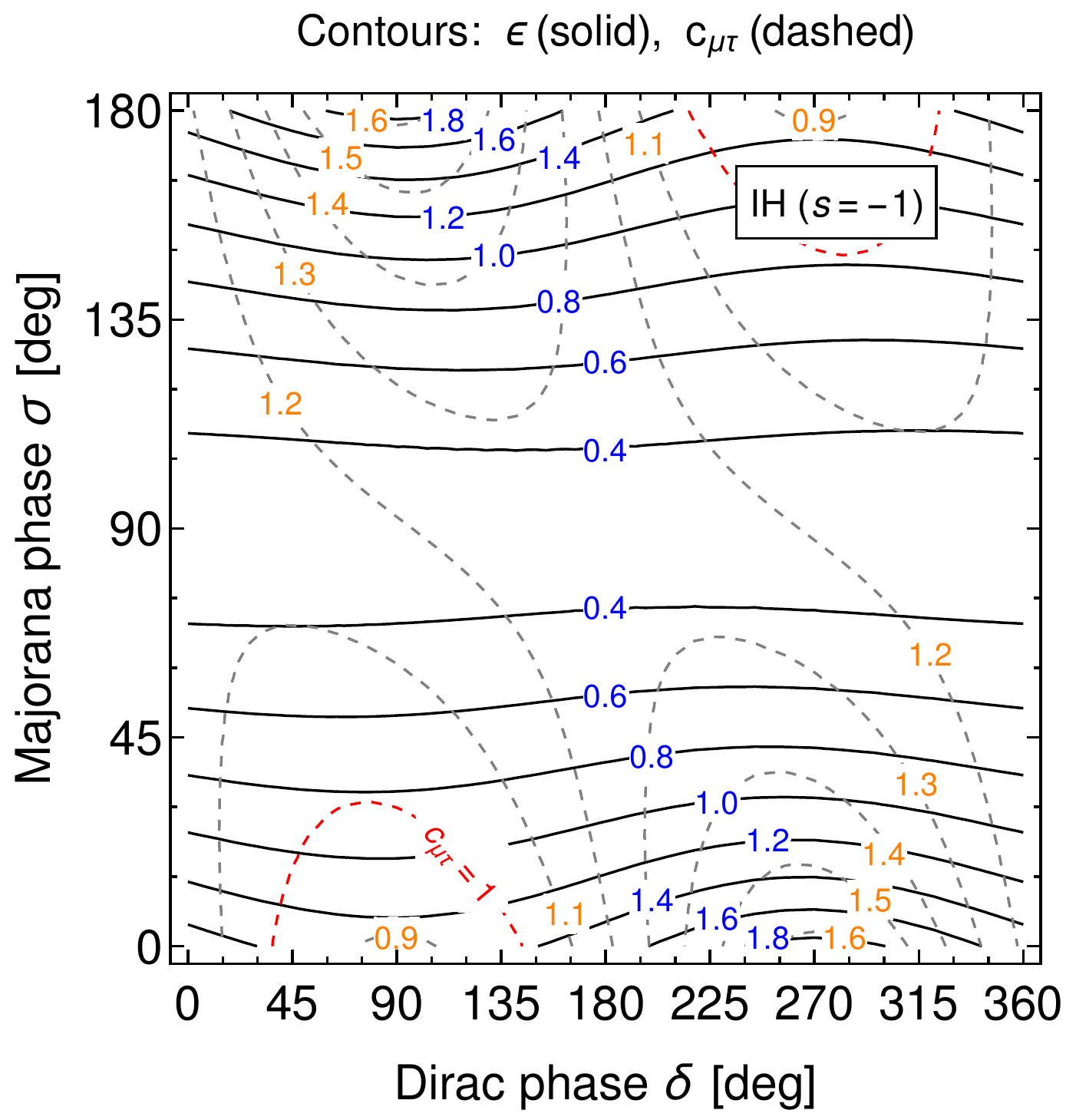}
\caption{Ratios of Yukawa couplings---$\left|y_{eI}\right|/\left|y_{\mu J}\right|$
(black solid contours, blue labels) and $\left|y_{\tau I}\right|/\left|y_{\mu J}\right|$
(gray dashed contours, orange labels)---in the flavor-aligned limit,
as functions of the $CP$-violating phases $\delta$ and $\sigma$.
Both ratios are shown for NH \textbf{(left panels)} and IH \textbf{(right panels)}
as well as for both possible signs of the parameter $z_I$ in the CIP,
$\textrm{sign}\left\{z_I\right\} = +1$ \textbf{(upper panels)} and 
$\textrm{sign}\left\{z_I\right\} = -1$ \textbf{(lower panels)}.
All ratios are calculated according to Eq.~\eqref{eq:ratios}, with all observables
except for $\delta$ and $\sigma$ being set to their respective best-fit
values, see Tab.~\ref{tab:data}.
In our flavor model,  $\left|y_{eI}\right|/\left|y_{\mu J}\right|$ and
$\left|y_{\tau I}\right|/\left|y_{\mu J}\right|$ are controlled by the
parameters $\epsilon$ and $c_{\mu\tau}$, respectively, see Eq.~\eqref{eq:yec}.
In this sense, the above plots may also be understood as predictions
for $\delta$ and $\sigma$ in dependence of $\epsilon$ and $c_{\mu\tau}$.
The expected values for $\epsilon$ and $c_{\mu\tau}$
correspond to $\epsilon \simeq \epsilon_0 \simeq 0.17$ (green solid line)
and $c_{\mu\tau} \simeq 1$ (red dashed line).
Demanding that $\epsilon \simeq 0.17$ and $c_{\mu\tau} \simeq 1$ simultaneously
leads to, in total, four different solutions (red dots in the lower-left panel).
This illustrates that our model predicts NH as well as a maximal Dirac phase,
$\delta \simeq \pm\pi/2$.}
\label{fig:ratios} 
\end{center}
\end{figure}


One of the main predictions of our Froggatt-Nielsen flavor model
is that the Yukawa couplings of the electron, muon, and tau flavors
have to exhibit a certain, characteristic hierarchy,
\begin{align}
\frac{\left|y_{e1}\right|^2 + \left|y_{e2}\right|^2}
{\left|y_{\mu 1}\right|^2 + \left|y_{\mu 2}\right|^2} = \epsilon^2 \,, \quad 
\frac{\left|y_{\tau1}\right|^2 + \left|y_{\tau2}\right|^2}
{\left|y_{\mu 1}\right|^2 + \left|y_{\mu 2}\right|^2} = c_{\mu \tau}^2 \,.
\label{eq:yec2}
\end{align}
Moreover, thanks to our approximate exchange symmetry,
$\left|y_{\alpha1}\right| \approx \left|y_{\alpha2}\right|$, this readily implies
\begin{align}
\frac{\left|y_{eI}\right|}{\left|y_{\mu J}\right|} \simeq \epsilon \,, \quad
\frac{\left|y_{\tau I}\right|}{\left|y_{\mu J}\right|} \simeq c_{\mu\tau} \,, \quad
I,J = 1,2 \,.
\label{eq:yec}
\end{align}
To check the compatibility of these predictions with the experimental data,
we shall now utilize the CIP in Eq.~\eqref{eq:CI} and examine the parameter
dependence of $\left|y_{eI}\right|/\left|y_{\mu J}\right|$ and
$\left|y_{\tau I}\right|/\left|y_{\mu J}\right|$ in the large-$\left|z_I\right|$
limit.
Making use of the results obtained in the previous section 
(see Eq.~\eqref{eq:alignment}), we immediately see that,
in this limit, both ratios may be expressed as follows
(using $M_1 \simeq M_2$),
\begin{align}
\left|z_I\right| \gg 1 \quad\Rightarrow\quad 
\frac{\left|y_{e I}\right|}{\left|y_{\mu J}\right|} \simeq
\left|\frac{V_e^s}{V_\mu^s}\right| \,,\quad 
\frac{\left|y_{\tau I}\right|}{\left|y_{\mu J}\right|} \simeq
\left|\frac{V_\tau^s}{V_\mu^s}\right| \,, \quad s = \textrm{sign}\left\{z_I\right\} \,.
\label{eq:ratios}
\end{align}
Here, recall that the $V_\alpha^\pm$ are defined as linear combinations
of elements ($V_{\alpha k}$ and $V_{\alpha l}$) of the rescaled PMNS matrix $V$,
see Eq.~\eqref{eq:defs}.
For our purposes, the message from Eq.~\eqref{eq:ratios} is that,
in the large-$\left|z_I\right|$ limit, all Yukawa ratios become
independent of the auxiliary parameter $z$.
On the assumption of flavor alignment, the ratios
$\left|y_{eI}\right|/\left|y_{\mu J}\right|$ and
$\left|y_{\tau I}\right|/\left|y_{\mu J}\right|$, therefore,
end up being functions of the low-energy observables encoded in $V$ only.
More explicitly, these observables consist of: the light-neutrino
mass eigenvalues $m_i$, the three neutrino mixing angles $\theta_{ij}$
as well as the $CP$-violating phases $\delta$ and $\sigma$.
Setting the light-neutrino mass eigenvalues and mixing angles to their measured
values (see Tab.~\ref{tab:data}) thus turns 
the ratios
$\left|y_{eI}\right|/\left|y_{\mu J}\right|$ and
$\left|y_{\tau I}\right|/\left|y_{\mu J}\right|$ into functions of
the yet undetermined phases $\delta$ and $\sigma$.
In Fig.~\ref{fig:ratios}, we plot these functions for both
NH and IH as well as for both possible signs of $z_I$ in the 
large-$\left|z_I\right|$ limit, $s = \textrm{sign}\left\{z_I\right\} = \pm 1$.
Our numerical results in Fig.~\ref{fig:ratios} lead us to several interesting
observations:
\begin{enumerate}
\item In the case of an inverted hierarchy, the ratio of the electron
and muon Yukawa couplings is bounded from below, 
$\left|y_{eI}\right|/\left|y_{\mu J}\right| \gtrsim 0.30$.
This is inconsistent with the expectation that, in our Froggatt-Nielsen
flavor model, we should rather find 
$\left|y_{eI}\right|/\left|y_{\mu J}\right| \simeq \epsilon \simeq \epsilon_0 \simeq 0.17$.
We therefore conclude that the IH case is unviable from the perspective
of our model.
\item The NH case remains, by contrast, feasible.
Along the green solid contours in the upper-left and lower-left panels
of Fig.~\ref{fig:ratios}, we find $\left|y_{eI}\right|/\left|y_{\mu J}\right| \simeq 0.17$
in agreement with the expectation from our flavor model.
This is a remarkable result for two reasons:
First of all, it tells us that our model clearly predicts a normal light-neutrino
mass ordering!
Second, this finding implies that, within the minimal seesaw model,
we are actually able to realize an approximate two-zero texture
in the neutrino Yukawa matrix.
This is noteworthy insofar as the requirement of
an \textit{exact} two-zero texture in the minimal seesaw model
can be shown to be inconsistent with the assumption of a normal
mass hierarchy~\cite{Harigaya:2012bw,Zhang:2015tea}.
When attempting to realize an \textit{exact} texture
(such as, e.g., $y_{e1} = y_{\mu2} = 0$),
one is unavoidably led to assume an inverted hierarchy.
As shown by the systematic study in~\cite{Rink:2016vvl},
the NH case only remains viable as long as one
allows for perturbations around the exact zeros.
\item As evident from the green contours in Fig.~\ref{fig:ratios},
our model is consistent with a broad range of $\left(\delta,\sigma\right)$
values.
Varying the parameter $c_{\mu\tau}$ between $1/2$ and $2$ (see the gray
dashed contours), we are able to realize all possible values of
the Majorana phase $\sigma$ and almost all possible values of the Dirac
phase $\delta$.
Meanwhile, the allowed values of $\delta$ and $\sigma$ are strongly
correlated with each other, as indicated by the green contours.
\item We do obtain more precise predictions for $\delta$ and $\sigma$
as soon as we impose more restrictive conditions on the parameter $c_{\mu\tau}$.
In general, $c_{\mu\tau}$ is expected to any number be of $\mathcal{O}\left(1\right)$.
However, if we insist on $c_{\mu\tau}$ being very close to unity,
$c_{\mu\tau} \simeq 1$ (see the red dashed contours),
only four viable combinations of $\delta$ and $\sigma$ remain
(see the red dots in the lower-left panel),
\begin{align}
\frac{\left(\delta,\sigma\right)}{\textrm{deg}} \simeq
\begin{cases}
\left(234, 7\right) & \textrm{(solution I)}\\                                  
\left(291, 9\right) & \textrm{(solution II)}\\   
\left(69, 171\right) & \textrm{(solution III)}\\   
\left(126,173\right) & \textrm{(solution IV)}
\end{cases} \,.
\label{eq:solutions}
\end{align}
All of these solutions feature $\delta$ values that imply close-to-maximal $CP$
violation, i.e., values either close to $90\,\textrm{deg}$ or $270\,\textrm{deg}$.
On top of that, solutions I and II are remarkably close to the current best-fit value
for the Dirac phase $\delta$ in the NH case, $\delta \simeq 261\,\textrm{deg}$,
see Tab.~\ref{tab:data}.
\end{enumerate}


Our last result (\#4) relies on the assumption that $c_{\mu\tau} \simeq 1$,
which may also be supported by several physical considerations.
First of all, $c_{\mu\tau} \simeq 1$ should be regarded as the naive
expectation from our flavor model.
In addition to that, setting $c_{\mu\tau}$ to a value close to unity
leads to an approximate mu-tau symmetry in the seesaw Lagrangian
(see \cite{Xing:2015fdg} for a review), which nicely complies with our FN
charge assignment.
Last but not least, it removes the last
free parameter in our model that we do not have theoretical control over.
With $c_{\mu\tau} \simeq 1$, Eq.~\eqref{eq:Lmodel} turns into
\begin{align}
\mathcal{L}_{\rm seesaw} \sim
- y_0 \left(\epsilon_0\,\ell_e + \ell_\mu + \ell_\tau\right)
\left(N_1 + i\,N_2\right) H
- \frac{1}{2}\,M \left(N_1 N_1 + N_2 N_2 \right) + \textrm{h.c.} \,,
\label{eq:Lmutau}
\end{align}
where we have also set $\epsilon$ to $\epsilon_0$ and $s = \textrm{sign}\left\{z_I\right\}$
to $s = -1$ according to our findings in Fig.~\ref{fig:ratios}.


\begin{figure}
\begin{center}
\includegraphics[width=0.67\textwidth]{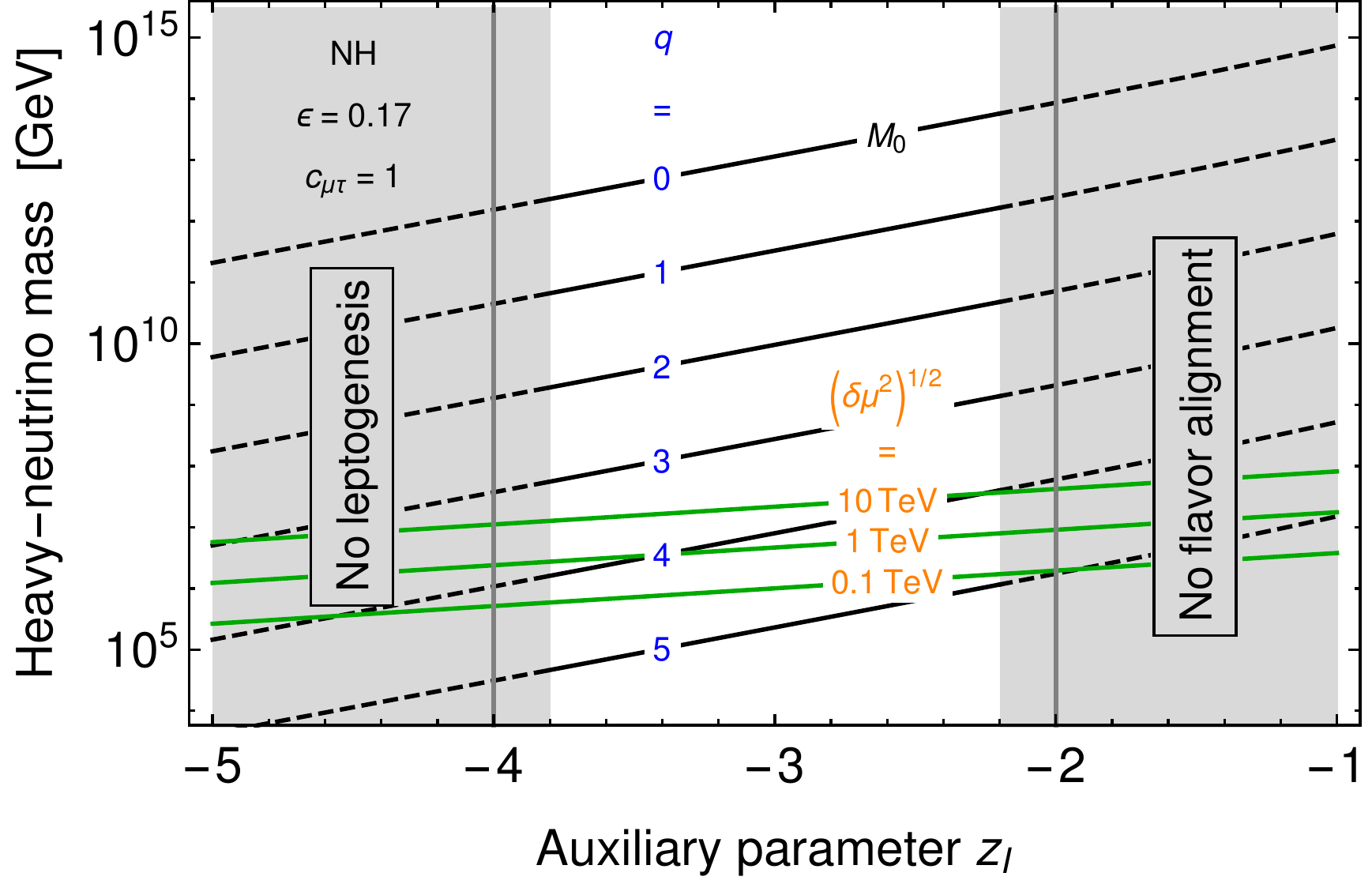}
\caption{Heavy-neutrino (pseudo-) Dirac mass $M$ as a function of
$q$ and $z_I$ (black lines), see Eq.~\eqref{eq:MqzI}.
We focus on $z_I$ values in the range $-4\lesssim z_I \lesssim -2$.
For larger values of $z_I$ (i.e., smaller $\left|z_I\right|$),
the condition of flavor alignment is no longer satisfied, see Fig.~\ref{fig:tan}.
Smaller values of $z_I$ (i.e., larger $\left|z_I\right|$), on the other hand,
do not allow for successful leptogenesis, see the discussion in~\cite{Bambhaniya:2016rbb}.
We also show the upper bound on $M$ deduced from the requirement of electroweak
naturalness, see Eq.~\eqref{eq:Mmax}, for three representative values of
$\delta\mu_{\rm max}^2$ (green lines).}
\label{fig:mass} 
\end{center}
\end{figure}


Thanks to its additional mu-tau symmetry, the
Lagrangian in Eq.~\eqref{eq:Lmutau} is even more minimal
than the one in Eq.~\eqref{eq:Lmodel}.
In fact, the only remaining parameters in Eq.~\eqref{eq:Lmutau} that are not fixed
by our flavor model are the FN charge $q$ as well as the mass scale $M_0$,
see Eq.~\eqref{eq:yM}.
Here, $M_0$ exhibits in particular a one-to-one correspondence to the parameter $z_I$.
In the flavor-aligned limit and for $\textrm{sign}\left\{z_I\right\} = -1$,
it follows from the relations in Eqs.~\eqref{eq:yM}, \eqref{eq:defs}, \eqref{eq:alignment}, 
and \eqref{eq:yec} that
\begin{align}
M_0\,e^{-2\,z_I} 
\simeq 2\,\epsilon_0^2 \left|V_e^-\right|^{-2} v_{\rm ew}
\simeq 2\, \left|V_\mu^-\right|^{-2} v_{\rm ew}
\simeq 2\, \left|V_\tau^-\right|^{-2} v_{\rm ew}
\simeq 4.6\times 10^{15}\,\textrm{GeV} \,.
\end{align}
This allows us to express the heavy-neutrino (pseudo-) Dirac mass
$M$ as a function of $q$ and $z_I$,
\begin{align}
M \simeq 4.6\times 10^{15}\,\textrm{GeV}\,\epsilon_0^{2q}\,e^{2\,z_I} \,.
\label{eq:MqzI}
\end{align}
We plot this function for several values of $q$ and for a representative
range of $z_I$ values in Fig.~\ref{fig:mass}.
In addition, we also show upper bounds on $M$ from the requirement
of electroweak naturalness, following the discussion in~\cite{Bambhaniya:2016rbb}.
The philosophy behind these bounds is the following:
Because of their Yukawa interactions with the SM leptons as well
as with the SM Higgs, the heavy
neutrinos yield radiative corrections, $\delta\mu^2$, to the mass-squared
parameter in the SM Higgs potential, $\mu^2$,
\begin{align}
\delta\mu^2 \approx \frac{M^3}{4\pi^2\,v_{\rm ew}^2} \cosh\left(2\,z_I\right)\sum m_i \,.
\end{align}
The larger these corrections are (compared to $v_{\rm ew}$), the more fine-tuning
is needed to eventually arrive at a SM Higgs mass of $125\,\textrm{GeV}$.
The requirement of electroweak naturalness, thus, imposes an upper bound on 
$\delta\mu^2$, which, in turn, translates into an upper bound on $M$,
\begin{align}
M \lesssim M_{\rm max} \simeq \left[\left(\cosh\left(2\,z_I\right)\sum m_i\right)^{-1}
4\pi^2\,v_{\rm ew}^2\,\delta\mu_{\rm max}^2\right]^{1/3} \,.
\label{eq:Mmax}
\end{align}
As evident from Fig.~\ref{fig:mass}, this bound, if taken seriously,
rules out FN charges smaller than 4.
Meanwhile, a heavy-neutrino FN charge of $q=4$ is marginally consistent
with the requirement of electroweak naturalness.
All larger FN charges, $q=5,6,\cdots$ lead to negligibly small
corrections to the Higgs mass parameter,
$\delta\mu^2 \lesssim \left(100\,\textrm{GeV}\right)^2$ and are, thus, perfectly allowed.


The authors of~\cite{Bambhaniya:2016rbb} also study resonant
leptogenesis~\cite{Flanz:1996fb,Pilaftsis:1997jf,Pilaftsis:2003gt}
in the minimal seesaw model.
They arrive at the conclusion that successful leptogenesis is, indeed,
feasible, provided that $\left|z_I\right|$ does not take too large a value,
$\left|z_I\right| \lesssim 4$ for $M \gtrsim 1\,\textrm{TeV}$.
Moreover, the authors of~\cite{Bambhaniya:2016rbb} conclude that the
requirements of successful leptogenesis and electroweak naturalness provide
the strongest constraints on the parameter space of the minimal seesaw model
for heavy-neutrino masses larger than about $10^4\,\textrm{GeV}$.
In this mass range, the bounds from vacuum (meta-) stability,
perturbativity, and lepton flavor violation turn out to be less constraining.
For our purposes, this means that---as can be seen from Fig.~\ref{fig:mass}---our
model does admit parameter solutions that comply with all of these theoretical
constraints and that, at the same time, still allow for successful leptogenesis!
All we have to do is to assign the heavy neutrinos an FN charge larger than $3$
and set the heavy-neutrino mass scale $M_0$ to a value of
$\mathcal{O}\left(10^{12}\cdots10^{14}\right)\,\textrm{GeV}$.


\subsection*{Conclusions and outlook}


In closing, let us summarize our \textit{most minimal}
realization of the type-I seesaw model.
The starting point of our construction was the minimal seesaw
model featuring only two rather than three heavy
neutrinos~\cite{Smirnov:1993af,King:1998jw,Frampton:2002qc,Ibarra:2003up}.
In two steps, we first embedded this model into the Froggatt-Nielsen model
of~\cite{Buchmuller:1998zf} and then made the assumption of an approximate
exchange symmetry that manifests itself as $N_1 \leftrightarrow i\,N_2$
in the heavy-neutrino Yukawa terms and as $N_1 \leftrightarrow N_2$
in the heavy-neutrino mass terms.
After eliminating the last undetermined parameter, $c_{\mu\tau}$,
by assuming an approximate mu-tau symmetry~\cite{Xing:2015fdg},
we then arrived at the following Lagrangian,
\begin{align}
\boxed{
\mathcal{L}_{\rm seesaw} \sim
- \epsilon_0^q \left(\epsilon_0\,\ell_e + \ell_\mu + \ell_\tau\right)
\left(N_1 + i\,N_2\right) H
- \frac{1}{2}\,\epsilon_0^{2q}\,M_0 \left(N_1 N_1 + N_2 N_2 \right) + \textrm{h.c.} \,,}
\label{eq:Lfinal}
\end{align}
modulo small corrections.
Here, $\epsilon_0$ denotes the FN hierarchy parameter.
Its value, $\epsilon_0 \simeq 0.17$, follows from demanding consistency
with the SM quark and charged-lepton mass spectra.
To obtain light-neutrino mass eigenvalues of the correct order of magnitude,
the heavy-neutrino mass scale $M_0$ needs to take a value of
$\mathcal{O}\left(10^{12}\cdots10^{14}\right)\,\textrm{GeV}$.
If one decides to ignore arguments concerning the naturalness of the electroweak
scale, the FN charge $q$ remains unconstrained, $q=0,1,2,\cdots$.
On the other hand, imposing the requirement of electroweak naturalness, one is
led to conclude that $q$ is bounded from below, $q\geq 4$.
However, even in this case, our model remains perfectly viable.
It complies with all theoretical constraints from naturalness,
vacuum (meta-) stability, perturbativity, and lepton flavor violation,
and still manages to account for the observed baryon asymmetry of
the universe via resonant leptogenesis.
This success of our model renders it an important benchmark scenario
for future experimental updates.


The Lagrangian in Eq.~\eqref{eq:Lfinal} is minimal for three reasons:
(i) instead of the usual three right-handed neutrinos, it only contains one pair
of pseudo-Dirac neutrinos;
(ii) it features flavor alignment in the neutrino Yukawa matrix, such that the
SM charged-lepton flavors basically couple to only one linear combination of
the heavy neutrino fields, $N_1 + i\,N_2$; and 
(iii) the neutrino Yukawa matrix exhibits an approximate two-zero texture,
which is the most minimal form one can hope to achieve in the NH case.
At this point, recall that our initial Lagrangian in Eq.~\eqref{eq:Lseesaw}
came with 11 degrees of freedom (DOFs) at high energies.
Out of these, we managed to eliminate a total of six:
two DOFs by our FN constraints on the neutrino
Yukawa couplings, see Eq.~\eqref{eq:yec2};
one DOF by requiring an approximate degeneracy
among the heavy-neutrino masses, see Eq.~\eqref{eq:exchange};
one DOF by our FN estimate of the pseudo-Dirac mass $M$, see Eq.~\eqref{eq:MqzI};
and two DOFs (i.e., the auxiliary parameter $z$) by our requirement
of flavor alignment in the neutrino Yukawa matrix, see Eq.~\eqref{eq:ratios}.
This leaves us with five DOFs---which can, however, be completely fixed
by the experimental data on the two mass-squared differences and three mixing angles.
In consequence, our model provides little to none parametric freedom.
In particular, it predicts a normal mass hierarchy
and a close-to-maximal Dirac phase, $\delta \simeq \pm \pi/2$, see
Eq.~\eqref{eq:solutions}.
Remarkably enough, this is in agreement with the latest hints
for $\delta$ values close to $\delta \simeq -\pi/2$, see Tab.~\ref{tab:data}.


Up to this point, we have been working in a field basis in which
the heavy-neutrino mass matrix is diagonal, see Eq.~\eqref{eq:Lseesaw}.
But it is also instructive to present the Lagrangian in Eq.~\eqref{eq:Lfinal}
in a slightly different basis.
Rotating the neutrino fields $N_1$ and $N_2$ to a new basis, such that
\begin{align}
\tilde{N}_1 = \frac{1}{\sqrt{2}} \left(N_1 + i\,N_2\right) \,, \quad
\tilde{N}_2 = \frac{1}{\sqrt{2}} \left(N_1 - i\,N_2\right) \,,
\end{align}
we clearly see that the heavy neutrinos form a pair of pseudo-Dirac
fermions with mass $\epsilon_0^{2q}M_0$,
\begin{align}
\mathcal{L}_{\rm seesaw} \sim
- \sqrt{2}\,\epsilon_0^q \left(\epsilon_0\,\ell_e + \ell_\mu + \ell_\tau\right)
\tilde{N}_1\, H - \epsilon_0^{2q}\, M_0\, \tilde{N}_1 \tilde{N}_2
+ \textrm{h.c.} \,.
\label{eq:LDirac}
\end{align}
Here, note that our exchange symmetry defined in terms of $N_{1,2}$
now acts as follows on $\tilde{N}_{1,2}$,
\begin{align}
N_1 & \leftrightarrow i\,N_2 \quad\Leftrightarrow\quad 
\tilde{N}_{1,2} \leftrightarrow \pm\phantom{i}\, \tilde{N}_{1,2} \,,
\label{eq:symmetries}\\ \nonumber
N_1 & \leftrightarrow \phantom{i}\,N_2 \quad\Leftrightarrow\quad 
\tilde{N}_{1,2} \leftrightarrow \pm i\, \tilde{N}_{2,1} \,.
\end{align}
Again, the first discrete symmetry plays the role of an approximate symmetry
of the Yukawa interactions, while the second discrete symmetry is an approximate
symmetry of the mass term.
Interestingly enough, the Lagrangian in Eq.~\eqref{eq:LDirac} is very similar
to the one appearing in the recently proposed model of
scalar neutrino inflation in supersymmetry~\cite{Nakayama:2016gvg}
(see also~\cite{Bjorkeroth:2016qsk,Kallosh:2016sej}).
In this model, the supersymmetric version of Eq.~\eqref{eq:LDirac} is
responsible for the exponential expansion of the universe during
inflation in the early universe.
The scalar superpartner of $\tilde{N}_2$ plays the role of the inflaton
field, while the scalar superparter of $\tilde{N_1}$ acts as a so-called
stabilizer field.
To realize successful inflation in this model, it is important that the Yukawa couplings
of the superfield $\tilde{N}_2$ be suppressed compared to the Yukawa couplings
of the superfield $\tilde{N}_1$.
In~\cite{Nakayama:2016gvg}, this requirement is met by assuming an approximate
shift symmetry in the direction of the inflaton field.
In our case, the Yukawa interactions of $\tilde{N}_2$ are, by contrast,
suppressed in consequence of the approximate discrete symmetries in Eq.~\eqref{eq:symmetries}.
In this sense, our model strongly parallels the construction in~\cite{Nakayama:2016gvg},
which may point to a deeper connection.
It thus appears worthwhile to embed our model into supersymmetry and 
study its implications for inflation more carefully.


In view of Eq.~\eqref{eq:LDirac}, it is also important
to note that our seesaw Lagrangian in Eqs.~\eqref{eq:Lfinal} and 
\eqref{eq:LDirac} only represents a leading-order approximation
that receives symmetry-breaking corrections.
The Lagrangian in Eq.~\eqref{eq:LDirac} exhibits, e.g., a lepton
number symmetry under which $\tilde{N}_1$ and $\tilde{N}_2$ carry charges
$-1$ and $+1$, respectively.
This indicates that Eq.~\eqref{eq:LDirac} predicts vanishing light-neutrino
masses, which is, of course, in conflict with the experimental data.
Nonzero neutrino masses follow in our model from the small symmetry-breaking
corrections to Eqs.~\eqref{eq:Lfinal} and \eqref{eq:LDirac}.
In our numerical analysis, we are correspondingly unable to \textit{predict}
the light-neutrino masses.
Instead, we use them as input data to calculate the elements of the rescaled
PMNS matrix $V$. 
By making use of the fact that the matrix $V$ determines the ratios of Yukawa couplings
in the flavor-aligned limit (see Eq.~\eqref{eq:ratios}), we are then able to
predict the $CP$ phases $\delta$ and $\sigma$, see Fig.~\ref{fig:ratios}.


In this Letter, we merely outlined the broad characteristics
of our model and more work is needed.
For one thing, our numerical analysis should be complemented by an analytical treatment.
This would help gain a better understanding of our numerical results.
In particular, one would like to achieve a better understanding
of the parameter dependence of our results in Fig.~\ref{fig:ratios}.
Such an analysis would allow to study the stability
of our findings under variations of the experimental input data. 
For another thing, one would like to embed the Lagrangian in Eq.~\eqref{eq:Lfinal}
into a larger framework for physics beyond the Standard Model.
The origin of the scale $M_0$ is, e.g., unaccounted for in our model.
It is therefore desirable to study possible connections between $M_0$ and, say,
the scale of grand unification (which is only four to two orders of magnitude
larger than $M_0$).
Moreover, one should attempt to explain the origin of the particular symmetries  
in Eq.~\eqref{eq:symmetries}. 
Such symmetries may, e.g., be related to certain boundary conditions
in extra-dimensional orbifold
constructions~\cite{Buchmuller:2004eg,Kobayashi:2008ih,Hall:2001rz,Hebecker:2002re}.
All of these questions are, however, beyond the scope of this paper,
which is why we leave them for future work.
For the time being, we content ourselves with the Lagrangian in Eq.~\eqref{eq:Lfinal},
concluding that it represents a remarkably simple realization of the
type-I seesaw mechanism.
Our model complies with all theoretical constraints and makes
important predictions for the neutrino mass ordering as well as the Dirac phase $\delta$.


\subsubsection*{Acknowledgements}


This work was supported by Grants-in-Aid for Scientific
Research from the Ministry of Education, Culture, Sports, Science and Technology (MEXT),
Japan, No.\ 26104009 (T.\,T.\,Y.), No.\ 26287039 (T.\,T.\,Y.) and No.\ 16H02176 (T.\,T.\,Y.)
as well as by the World Premier International Research Center  Initiative
(WPI Initiative), MEXT, Japan (T.\,T.\,Y.).



\end{document}